\documentclass[conference]{IEEEtran}

\IEEEoverridecommandlockouts

\usepackage{cite}
\usepackage{amsmath,amssymb,amsfonts}
\usepackage{graphicx}
\usepackage{textcomp}
\usepackage{xcolor}

\usepackage{subcaption}

\usepackage{tikz}
\usepackage{pgfplots}
\pgfplotsset{compat=1.18}

\usepackage{cleveref}
\usepackage{csquotes}
\usepackage{booktabs}

\usepackage{flushend}

\usetikzlibrary{patterns}

\definecolor{tol1}{HTML}{33bbee}
\definecolor{tol2}{HTML}{009988}
\definecolor{tol3}{HTML}{ee7733}
\definecolor{tol4}{HTML}{cc3311}
\definecolor{tol5}{HTML}{66CCEE}
\definecolor{tol6}{HTML}{CD7F32}

\definecolor{plt1}{HTML}{4053d3}
\definecolor{plt2}{HTML}{ddb310}
\definecolor{plt3}{HTML}{b51d14}
\definecolor{plt4}{HTML}{00beff}
\definecolor{plt5}{HTML}{fb49b0}

\pgfplotscreateplotcyclelist{bwb}{%
  thick, tol1\\%
  thick, tol2\\%
  thick, tol3\\%
  thick, tol4\\%
  thick, tol5\\%
  thick, tol6\\%
}

\usepackage{acro}
\DeclareAcronym{cots}{
  short = COTS,
  long = off-the-shelf,
}

\newcommand\MP{\ensuremath{\mathrm{MP}}\xspace}
\newcommand\SNR{\ensuremath{\mathrm{SNR}}\xspace}

\newcommand\ours{\textsc{AdaJscc}\xspace}
\newcommand\baseline{Baseline\xspace}

\DeclareAcronym{tiff}{
  short = TIFF,
  long = tag image file format,
}

\DeclareAcronym{mse}{
  short = MSE,
  long = mean squared error,
}

\DeclareAcronym{awgn}{
  short = AWGN,
  long = additive white Gaussian noise,
}

\DeclareAcronym{pdf}{
  short = PDF,
  long = probability-density function
}

\DeclareAcronym{leo}{
  short = LEO,
  long = low Earth orbit,
}

\DeclareAcronym{ldpc}{
  short = LDPC,
  long = low-density parity-check,
}

\DeclareAcronym{jscc}{
  short = JSCC,
  long = joint source and channel coding,
}

\DeclareAcronym{djscc}{
  short = DJSCC,
  long = deep joint source and channel coding,
}

\DeclareAcronym{snr}{
  short = SNR,
  long = signal-to-noise ratio,
}

\DeclareAcronym{cnn}{
  short = CNN,
  long = convolutional neural network,
}

\DeclareAcronym{psnr}{
  short = PSNR,
  long = peak signal-to-noise ratio,
}

\DeclareAcronym{prelu}{
  short = PReLU,
  long = parameterized rectified linear unit,
}

\DeclareAcronym{relu}{
  short = ReLU,
  long = rectified linear unit,
}

\DeclareAcronym{jscc-sat}{
  short = \textsc{JSCC-Sat},
  long = {joint source-and-channel coding for small satellite applications}
}

\DeclareAcronym{los}{
  short = {LOS},
  long = {line of sight}
}

\DeclareAcronym{esa}{
  short=ESA,
  long=European Space Agency
}

\usepackage{xspace}

\newcommand\jpegtwok{JPEG\,2000\xspace}

\newcommand\sentinelii{Sentinel-2\xspace}

\begin{document}

\title{Adaptable Deep Joint Source-and-Channel Coding for Small Satellite Applications}

\author{\IEEEauthorblockN{Olga Kondrateva}
\IEEEauthorblockA{\textit{Technical University of Darmstadt}\\
Berlin, Germany \\
olga.kondrateva@kom.tu-darmstadt.de}
\and
\IEEEauthorblockN{Stefan Dietzel}
\IEEEauthorblockA{\textit{Merantix Momentum GmbH}\\
Berlin, Germany \\
stefan@merantix-momentum.com}
\and
\IEEEauthorblockN{Bj\"orn Scheuermann}
\IEEEauthorblockA{\textit{Technical University of Darmstadt}\\
Darmstadt, Germany \\
scheuermann@kom.tu-darmstadt.de}
}

\maketitle

\begin{abstract}
Earth observation with small satellites serves a wide range of relevant applications.
However, significant advances in sensor technology (e.g., higher resolution, multiple spectrums beyond visible light) in combination with challenging channel characteristics lead to a communication bottleneck 
when transmitting the collected data to Earth.
Recently, joint source coding, channel coding, and modulation based on neuronal networks has been proposed to combine image compression and communication.
Though this approach achieves promising results when applied to standard terrestrial channel models, 
it remains an open question whether it is suitable for the more complicated and quickly varying satellite communication channel. 
In this paper, we consider a detailed satellite channel model accounting for different shadowing conditions and train an encoder-decoder architecture with realistic \sentinelii satellite imagery.
In addition, to reduce the overhead associated with applying multiple neural networks for various channel states, 
we leverage attention modules and train a single adaptable neural network that covers a wide range of different channel conditions.
Our evaluation results show that the proposed approach achieves similar performance when compared to less space-efficient schemes that utilize separate neuronal networks for differing channel conditions.   
\end{abstract}

\acresetall
\section{Introduction}

Earth observation using sensor data acquired by satellites has gained more and more attention over the last years.
Common use cases include environment monitoring \cite{rs14030589}, disaster management \cite{barmpoutis2020}, and many more \cite{radix,MarCO}.
The increasing momentum can be explained by two major factors.
First, technological advances have allowed to build smaller satellites, which use more off-the-shelf components and can be deployed more easily.
A prime example are CubeSats, which operate in \ac{leo} and consist of $10 \times 10 \times 10$\,cm units \cite{cubesat2020}.
Second, sensor technology has improved greatly.
Besides higher spatial resolution, modern sensors support a wider spectrum range, exceeding that of visible light.
Hyperspectral images include infrared and other bands, which allow to monitor vegetation, clouds, and other phenomena.

These advances in satellite and sensor technology, however, are not met by equal improvements in communication capacity.
CubeSats and other small satellites have a constrained energy budget.
And, unlike geostationary satellites, they orbit the Earth several times per day with high speed, limiting the communication windows with ground stations.
Their high velocity and harsh channel conditions due to weather and low elevation angles further lead to high packet loss rates and complicate communication \cite{nogales2018}.

Therefore, efficient and robust communication systems are necessary to support demanding Earth observation applications.
Typically, source coding, channel coding, and modulation schemes are applied to convert image sensor data into physical layer channel symbols.
Source coding serves to compress sensor images, often in a lossy fashion (e.g., \jpegtwok) \cite{sentinel-2-user-handbook}.
Channel coding (e.g., \ac{ldpc}) is then used to enable error correction, counteracting packet loss due to the harsh channel conditions.
More recently, approaches that consider these coding mechanisms jointly have emerged, promising better performance than individual coding schemes \cite{6408177}.
Although Shannon's theory \cite{cover1991elements} states that separate optimization can yield optimal results, its assumptions, such as infinite code block lengths, are not always true in practice.

Neuronal networks have provided a feasible way to implement such a joint coding approach, improving over earlier work, which was too complex to be useful in practice \cite{9838671}.
Neuronal-network-based joint coding approaches have been proposed for both terrestrial communication \cite{Bourtsoulatze2019} and satellite applications \cite{satjscc}.
So far, a major limitation has been that the joint encoder and decoder has been trained based on fixed assumptions about channel characteristics, such as the \ac{snr} based on a simple \ac{awgn} channel model. 
To foster practical application of this approach, 
a more realistic channel model accounting for specific satellite channel conditions should used.
Another limitation concerns the adaptability of this approach to changing channel conditions. In the previous work, separate neuronal networks were trained independently under the assumption that they can be switched between by both the satellite and the ground station.
In this scenario, considering more complex channel models would easily lead to a combinatorial explosion of parameters, and consequently a prohibitive amount of separate neuronal networks.
Therefore, separate networks obviously restrict the total number of channel parameter combinations that can reasonably be covered.

In this paper, we propose a neuronal-network-based \ac{jscc} approach that uses a realistic satellite channel model to capture a wide range of channel conditions.
To manage the resulting complexity, we employ attention modules, which allow to parametrize models for different channel conditions rather than training separate models.
To characterize the channel, we use Fontán et al.'s model \cite{fontan2001} based on Markov chains.
It is applicable to non-geostationary small satellites and models a number of channel characteristics, such as multipath propagation and shadowing.
The model distinguishes multiple environments -- such as, urban, tree shadow, or open -- and three channel states -- \ac{los}, shadowing, and deep shadowing.
In addition, we also derive expected \ac{snr} values based on satellite elevation angles, distance, and other parameters.

Evidently, training separate networks for all possible environments, channel states, elevation angles, and so forth, would lead to a prohibitive large number of individual neuronal networks.
Therefore, we use a model architecture that is enriched with so-called attention modules, which are compact layers that can be added in between the neuronal network's other layers to learn and focus on particular input parameters.
Essentially, these modules allow to parametrize the network for different actual channel conditions.
Although the resulting network with attention modules is larger than one trained for a specific set of channel conditions, it is considerably smaller than considering a set of separate networks, one for each channel condition.

Our evaluation shows that our approach performs similar to separate, individual networks for different channel conditions while requiring significantly lower storage overhead.

Thus, our main contributions can be summarized as follows:
\begin{enumerate}
  \item We enhance \ac{jscc} with a realistic channel model for small satellite applications.
  \item We apply an attention-module-augmented neuronal network architecture to be able to use a single network for a wide range of realistic channel characteristics.
  \item We evaluate our approach using a set of realistic \sentinelii Earth observation data for different channel characteristics and compression ratios.
  \item We evaluate the behavior of our approach in the case of channel mismatch, i.e., when the actual channel conditions differ from the assumptions made by the sender and receiver.
\end{enumerate}

The remainder of this paper is organized as follows.
In \Cref{sec:related_work}, we review existing work on \ac{jscc} and channel models.
Next, we provide an overview of our system model in \Cref{sec:system_model} before explaining our mechanism in detail in \Cref{sec:our_approach}.
\Cref{sec:evaluation} details our evaluation results using \sentinelii mission data.
We conclude the paper in \Cref{sec:conclusion}.

\section{Related Work}
\label{sec:related_work}

Combining source and channel coding (\acs{jscc}) has been an active research topic over the last decades.
A number of works compare \ac{jscc} to separate coding and investigate theoretical bounds for the achievable rates \cite{gallager1968information,1614076,4557472}
Apart from theoretical studies, various practical \ac{jscc} schemes have been proposed.
Yu et al. use optimization techniques to jointly optimize source and channel coding parameters to minimize energy consumption~\cite{Wei2004}, and
Xu et al. consider distributed \ac{jscc} for video transmission~\cite{4205066}. 

Recently, deep-learning-based methods have been successfully applied for source coding as well as for channel coding. In particular, various studies show that deep source coding is able to outperform traditional source coding techniques, such as JPEG and JPEG2000~\cite{toderici2016variable, ballé2017endtoend, Hu2022}.
Similarly, in the domain of channel coding, O'Shea et al. have demonstrated that performance close to Hamming codes can be achieved using an encoder-decoder architecture~\cite{8054694}. 
Based on these results, the use of deep learning has been extended to \ac{jscc}.
One of the first encoder-decoder architectures combining source coding, channel coding and modulation has been proposed by Bourtsoulatze et al.~\cite{Bourtsoulatze2019} and was later improved by a number of works \cite{Xuan2021,Kurka2021,yang2021_2}.
Also, the use of attention modules has been proposed to dynamically adapt to different \ac{snr} levels~\cite{9438648}. 
Furthermore, Kondrateva et al. proposed a JSCC encoder-decoder architecture suitable for transmission of satellite images \cite{satjscc}.
In contrast to our approach, these works limit their consideration to simple channel models that do not take into account the specifics of satellite communication.

Although different communication technologies, such as VLC~\cite{nakajima2012} and laser~\cite{welle2018}, have been considered for small satellites, RF-based approaches remain predominant.
Therefore, we briefly review channel models for satellite-to-ground RF-based communication.   
Loo's model~\cite{1623307} combines log-normal and Rayleigh distributions to model the \ac{los} and multipath components, respectively.
Corazza-Vatalaro's model~\cite{Corazza1994ASM}, in contrast, uses the product of Rician and log-normal distributions to describe the multipath component. 
Various extensions for Loo's and Corazza-Vatalaro's models have been proposed~\cite{596315, 661055}.
All these models are static in the sense that they use a single distribution to describe all propagation conditions.
To model signal changes over time, dynamic models \cite{966585,4151152,7779114,8693582}, which are mostly based on Markov chains, can be used \cite{9079470}.
For our evaluation we chose Fontán et al.'s model \cite{966585}, as it was tested for a broad range of frequencies in different environments and considers different elevation angles.

\section{Earth Observation Missions}
\label{sec:system_model}

\begin{figure*}
  \begin{subfigure}{.3\linewidth}
    \includegraphics[width=\linewidth]{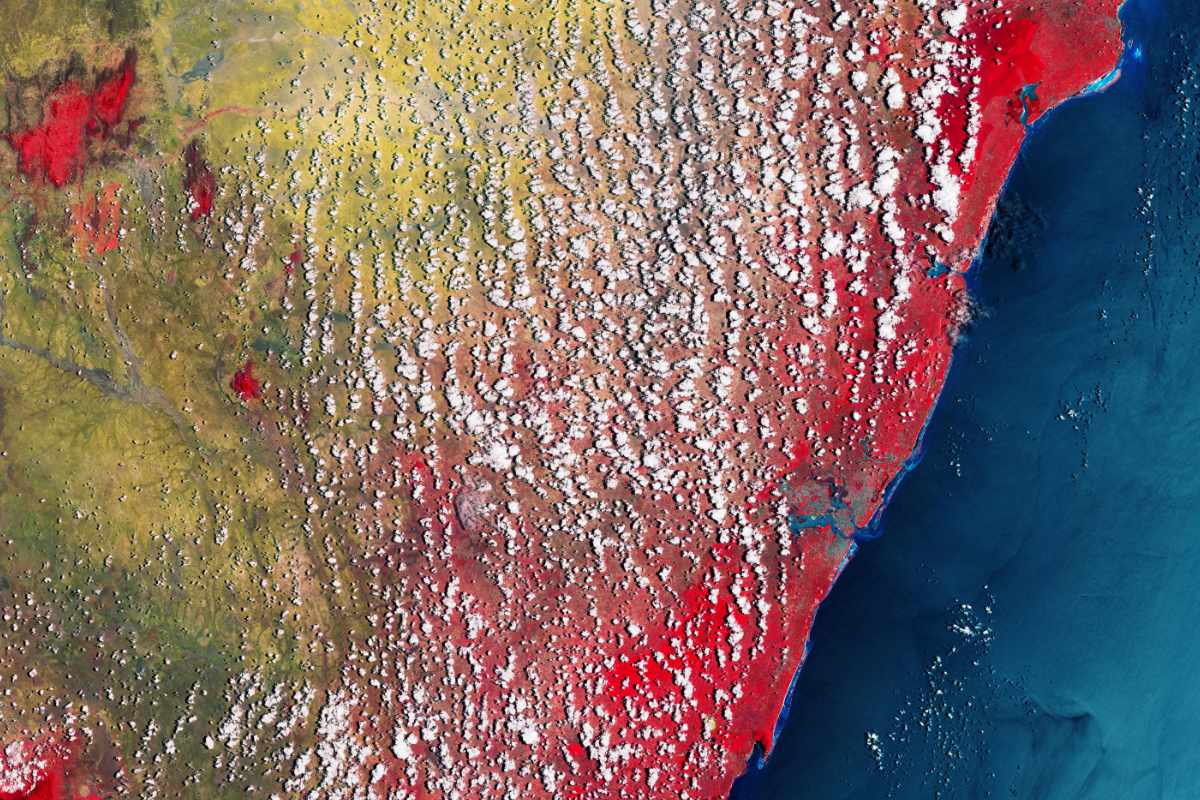}
    \caption{Southeast Kenya}
    \label{fig:sentinel_kenya}
  \end{subfigure}
  \hfill
  \begin{subfigure}{.3\linewidth}
    \includegraphics[width=\linewidth]{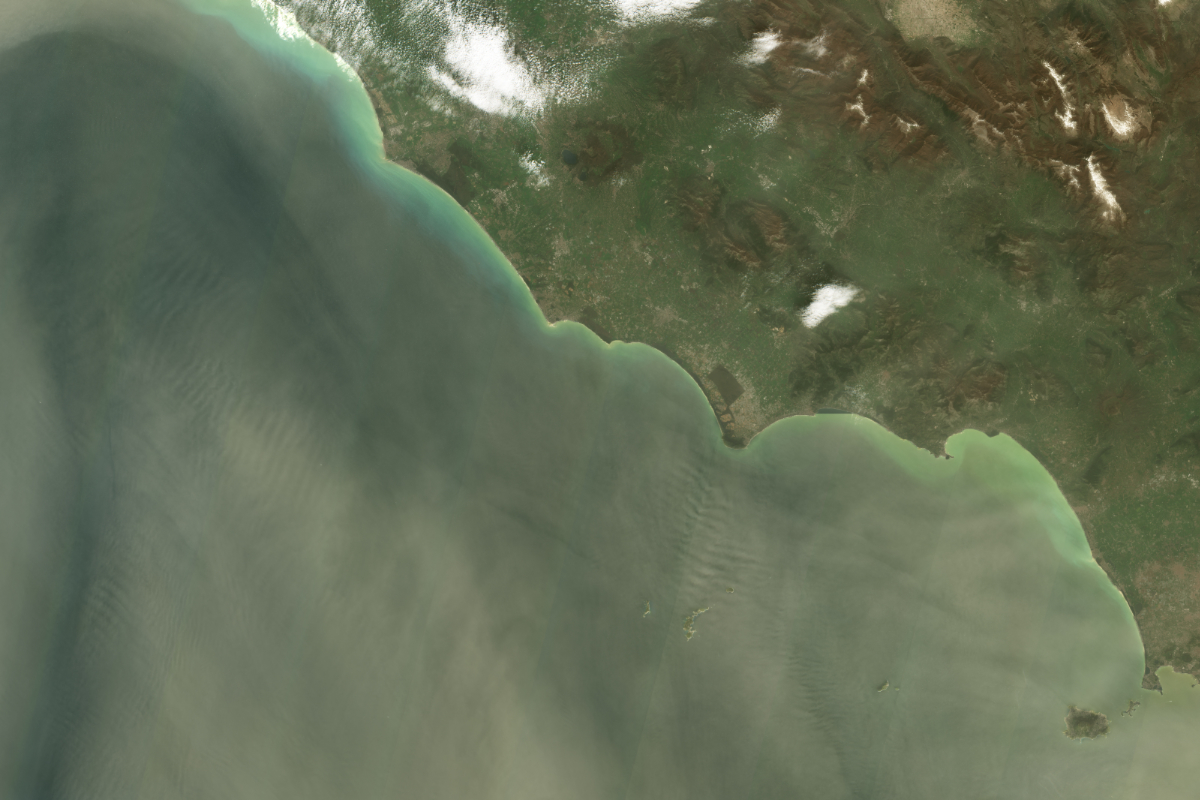}
    \caption{Sahara dust}
    \label{fig:sentinel_sahara}
  \end{subfigure}
  \hfill
  \begin{subfigure}{.3\linewidth}
    \includegraphics[width=\linewidth]{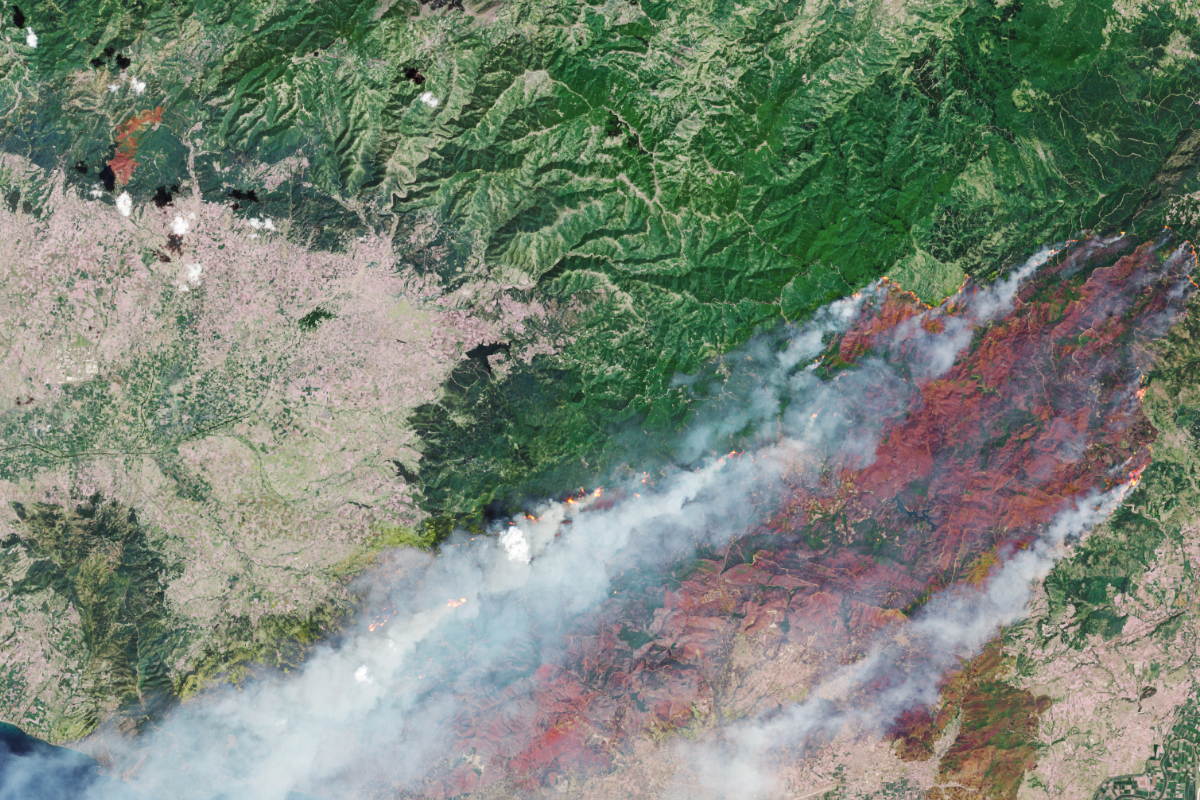}
    \caption{Wildfires in Greece}
    \label{fig:sentinel_greece}
  \end{subfigure}

  \caption{Example images from the \sentinelii mission. (Credit: processed by ESA, CC BY-SA 3.0 IGO)}
  \label{fig:sentinelii}
\end{figure*}

With recent advances in space and image sensing technologies, Earth observation using small satellites has gained more and more attention over the past years.
As an example use case for the methods we propose in this paper, we briefly introduce ESA's \sentinelii mission.
We also use a subset of the mission's dataset for our evaluation.
Moreover, we give an overview of how neuronal networks are used onboard satellites for image processing.

\sentinelii \cite{sentinel2} is operated by the \ac{esa} as part of the Copernicus program using \ac{leo} satellites.
The first satellites were launched in 2015, with more launches following in 2017 and 2024.
The missions comprise two satellites, which orbit the Earth such that each spot on the surface is revisited approximately every five days.
With their sensors, they cover a strip of land that is 290 km wide with each pass.
In addition to visible light sensors, the satellites are equipped with sensors that cover additional frequencies, such as infrared \cite{sentinel-2-user-handbook}, which allow to capture land use and vegetation.
In total, the satellite images comprise information in 12 different bands.
Moreover, the sensors feature relatively high optical resolutions between 10 and 60 meters.

\Cref{fig:sentinelii} shows three example use cases for Earth observation images taken from \ac{esa}'s homepage.
The first (\Cref{fig:sentinel_kenya}) is a false-color image of southeast Kenya, which was generated by overlaying \sentinelii's near-infrared channel onto the visual spectrum.
The bright red colors indicate higher plant density and health, as alive plants reflect near-infrared light.
Thereby, the dense vegetation in the coastal regions can easily be distinguished from the hinterland regions.
\Cref{fig:sentinel_sahara} shows a dust storm originating from the Sahara desert.
\sentinelii provides valuable insights for air pollution monitoring, and due to the short revisit time, storms can be monitored as they develop.
Finally, \Cref{fig:sentinel_greece} shows wildfires in Greece in 2023.
For the visualization, the shortwave infrared spectrum was merged with the visible light spectrum, showing the fire front.
Dark brown areas show the burned area.
Thereby, these images provided valuable insights for civil protection authorities.
By using more efficient and more robust image compression and transmission, sensor data in small satellite missions can be transmitted and used faster, allowing even more rapid responses.

In this paper, we propose a neuronal-network-based \ac{jscc} approach to improve communication.
As small satellites are severely power-constrained, necessary hardware resources need to be taken into account when considering the feasibility of our solution.
Recently, a number of processing platforms have been successfully evaluated  in space and small satellite scenarios.
Two examples are the Intel Movidius Myriad 2 and STM32 Microcontrollers, which have been used to identify stars and only consume approximately 1 Watt power \cite{8556744}.
During the $\Phi$-Sat mission, deployment of machine learning models has been evaluated using an Intel Movidius Myriad processor, as well.
Similarly, Nvidia's TX2 SoC, which is compatible with the CubeSat standard's power constraints, was able to detect cargo ships \cite{8556744}.
Finally, even relatively large standard machine learning models, such as VGG19 \cite{DBLP:journals/corr/SimonyanZ14a} and ResNet50 \cite{7780459}, have been evaluated on the International Space Station (ISS), operating on the Qualcomm Snapdragon 855 and Intel Movidius Myriad X processors \cite{9884906}.

\section{Adaptable \ac{jscc}}
\label{sec:our_approach}

The goal of our approach is to transmit sensor data $x \in \mathbb{R}$ acquired by small satellites in \ac{leo} to a ground station over a bandwidth-constrained channel.
The sensor data comprises optical images from multiple spectrums as defined in \Cref{sec:system_model}.
Assuming lossy compression and transmission, the ground station reconstructs an approximation $\hat{x}$ of the original data $x$ with the goal that $\hat{x}$ is as close to $x$ as possible.
Next, we give an overview of the protocol before we explain the neural network architecture using attention modules and the channel model in more detail.

\subsection{Architecture overview}

\begin{figure}
  \includegraphics[width=\linewidth]{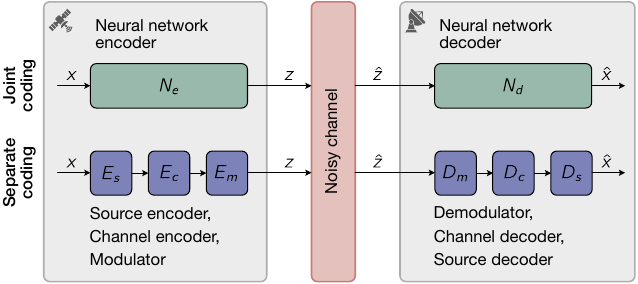}

  \caption{Overview of our communication architecture and comparison with traditional separate-encoder designs.}
  \label{fig:overview}
\end{figure}

To introduce our protocol's main components, we first explain the basic \ac{jscc} communication architecture and contrast it to the traditional case that uses separate encoders and decoders (\Cref{fig:overview}).
With separate coding (shown in the bottom half), the sensor data $x \in \mathbb{R}^n$ is first encoded using a source encoder, such as \jpegtwok.
Afterwards, a channel encoder (e.g., \ac{ldpc}) adds redundancy to the compressed signal in order to protect it against possible transmission errors.
The modulator component then translates the channel encoder's output to physical layer samples $z \in \mathbb{C}^k$, which can be transmitted over a noisy channel.
The ground station receives the distorted signal $\hat{z}$ and uses corresponding components in inverse order to decode an approximation of $x$.
The demodulator $D_m$ translates the samples back to bits, which serve as input to the channel decoder $D_c$.
The channel decoder reconstructs the compressed image data, correcting transmission errors as far as possible.
The source decoder $D_s$ finally approximately reconstructs the original optical sensor data as $\hat{x}$.

In the case of deep \ac{jscc} approach (upper half of \Cref{fig:overview}), in contrast, parts of a single neuronal network are used as encoder $N_e$ and decoder $N_d$ by the satellite and ground station, respectively.
These components jointly perform the source coding, channel coding, modulation, and their corresponding inverse operations.

We use an encoder-decoder neuronal network architecture based on \cite{satjscc} to implement $N_e$ and $N_d$:
During training, the encoder ($N_e$) and decoder ($N_d$) parts of the network are linked together using a realistic channel model for noisy satellite links.
To better reflect the channel conditions, we replace the \ac{awgn} channel used in \cite{satjscc} with a more realistic Fontán et al.'s statistical channel model~\cite{966585}.
The encoder-decoder model is then trained using an image dataset derived from the body of \sentinelii mission data.
Normally, a separate model would need to be trained for each characteristic set of channel conditions \cite{satjscc}.
In order to use a single network for a wide array of channel conditions, we augment the model with so-called attention modules \cite{wireless-attention-modules} as part of both the encoder and decoder during training.
These essentially allow to parameterize the network during later operation to cater to changing channel conditions.
The trained network is then separated into the encoder component $N_e$, which is used by the satellite and the decoder component $N_d$, which is used by the ground station.

\subsection{Encoder-decoder architecture with attention modules}

\begin{figure}
  \includegraphics[width=\linewidth]{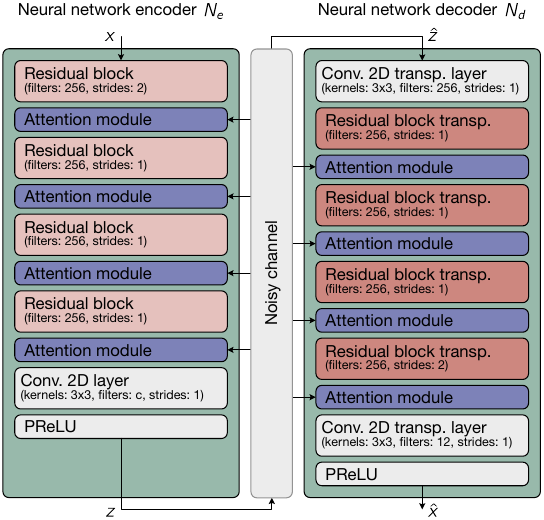}

  \caption{Network architecture overview.}
  \label{fig:architecture-overview}
\end{figure}

The encoder-decoder network architecture is shown in \Cref{fig:architecture-overview}.
As basis, we use the network structure proposed in \cite{satjscc}, which adapts ResNet \cite{resnet} to the \ac{jscc} use case.
To make the architecture more flexible, we add attention modules, as well as a more realistic channel model.
The block on the left serves as neural network encoder, jointly performing source coding, channel coding, and modulation.
Compression is achieved by translating the input $x \in \mathbb{R}^n$ to channel symbols $z \in \mathbb{C}^k$, where $k < n$.
By adjusting $k$, the system's compression ratio $k/n$ can be defined.

\begin{figure*}
  \begin{subfigure}{.3\linewidth}
    \centering
    \includegraphics[height=10em]{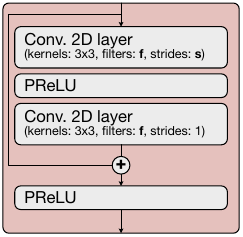}
    \caption{Residual block $(f, s)$}
    \label{fig:residual}
  \end{subfigure}
  \hfill
  \begin{subfigure}{.3\linewidth}
    \centering
    \includegraphics[height=10em]{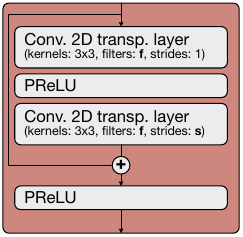}
    \caption{Residual block transpose $(f, s)$}
    \label{fig:residual-transpose}
  \end{subfigure}
  \hfill
  \begin{subfigure}{.3\linewidth}
    \centering
    \includegraphics[height=10em]{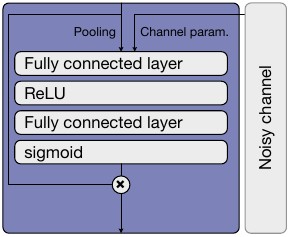}
    \caption{Attention module}
    \label{fig:attention}
  \end{subfigure}

  \caption{Specific block architectures used within the encoder-decoder network.}
  \label{fig:blocks}
\end{figure*}

For the encoder, we use four residual blocks with 256 filters and a kernel size of $3 \times 3$.
The structure of the residual blocks is shown in \Cref{fig:residual}.
Each residual block is followed by an attention module as proposed in \cite{wireless-attention-modules}.
\Cref{fig:attention} shows the structure of the attention modules.
These modules allow to alter the feature weights of their preceding residual blocks in order to accommodate different channel conditions.
To do so, a number of channel parameters are added as additional input to the attention modules.
Based on these parameters, a number of scaling parameters are learned that either increase or decrease the connection strength to the next module, depending on the specific channel condition.
The attention modules operate in three steps.
First, global average pooling is applied to the previous residual block's output to make available the global context information.
The pooled output is then concatenated with variables representing channel parameters.
Second, a simple neural network structure -- two fully connected layers plus \ac{relu} and sigmoid as activation functions -- is used to predict the proper scaling factors based on the channel condition parameters.
Finally, the previous residual layer's output is multiplied with the predicted scaling factors to implement the attention-based scaling.

As the last layer of the encoder component, we use a convolutional layer that contains $c$ filters.
The value for $c$ is chosen based on the desired compression ratio $k/n$.
Finally, \ac{prelu} is used as activation function.
The result is a vector $\tilde{z}$ comprising $k$ complex numbers.
The vector is normalized to enforce an average transmit power constraint $P$:
\begin{equation}
  z = \sqrt{kP}\frac{\tilde{z}}{\sqrt{\tilde{z}^{*} \tilde{z}}}
\end{equation}

The middle component of the architecture is a non-trainable channel layer, which models a realistic representation of the channel conditions in satellite communication.
For the channel, we use Fontán et al.'s statistical channel model~\cite{966585}, which assumes a log-normal distribution with mean $\alpha$ and standard deviation $\psi$ for the direct signal component and a Rayleigh-distributed multipath component with average power \MP. We also model Gaussian noise based on \SNR.
We present the channel model in more detail in \Cref{sub:channel_model}.

The block on the right serves as neural network decoder, which translates the potentially corrupted channel symbols $\hat{z}$ back to an approximation $\hat{x}$ of the original data.
The decoder follows the same architecture design: a convolutional transpose layer is followed by four residual transpose blocks plus corresponding attention modules,  another convolutional transpose layer, and a \ac{prelu} activation function.

During training, we use the average \ac{mse} as loss function:
\begin{equation}
  \mathcal{L} = \mathrm{MSE} = \frac{1}{N} \sum_{i=1}^{N} d \bigl(x_i, \hat{x}_i\bigr),
\end{equation}
where $N$ is the number of samples and $d(x, \hat{x}) = || x - \hat{x} ||^2$ is the \ac{mse} distortion.

To evaluate the quality of the reconstructed image signal, we use the \ac{psnr} metric:
\begin{equation}
  \mathrm{PSNR} = 10 \log_{10}\frac{\mathrm{MAX}^2}{\mathrm{MSE}},
\end{equation}
where $\mathrm{MAX}$ is the maximum possible pixel value.
The \ac{psnr} captures how much the original signal value is affected by distorting noise.

\subsection{Channel model for \ac{leo} satellites}
\label{sub:channel_model}

The channel model serves as non-trainable layer in between the encoder and decoder components of the neural network model.
We compute the channel output symbols $\hat{z}$ as follows:
\begin{equation}
  \hat{z} = zh + n,
\end{equation}
where $h$ is the channel gain, $n$ is Gaussian noise, and $z$ are the input symbols.
Next, we explain how $h$ and $n$ are calculated.

To account for specific satellite channel conditions, we compute $h$ using Fontán et al.'s statistical channel model~\cite{966585}. 
Since the satellite channel characteristics heavily depend on shadowing conditions, 
they generally cannot be accurately modeled using a single distribution.
Rather, the model introduces multiple states describing different degrees of shadowing. 
The overall \ac{pdf} is given as a sum of of the individual state \acp{pdf} multiplied by the probabilities of being in a given state.
For example, assuming a two-state model that distinguishes only between \ac{los} and shadow conditions: 
\begin{equation}
p_{\mathrm{overall}}(r) = x_{\mathrm{LOS}} \cdot p_{\mathrm{LOS}}(r)
  + x_{\mathrm{shadow}} \cdot p_{\mathrm{shadow}}(r),
\end{equation}
where $x_{\mathrm{LOS}}$ is the probability that the channel is in \ac{los} shadowing conditions, $p_{\mathrm{LOS}}$ is the \ac{pdf} of the amplitude variations $r$ under \ac{los} conditions, $x_{\mathrm{shadow}}$ is the probability that the channel being in shadow conditions, and $p_{\mathrm{shadow}}$ is the \ac{pdf} of the signal variations under shadow conditions.

To capture state durations, Markov chain models are typically introduced.
That is, the propagation channel at a given time is described by the Markov chain's state probability matrix and the state transition probability matrix.
The former contains the probabilities of being in a particular state, and the latter the probability of changing from one state to another.

Specifically, Fontán et al.'s channel model \cite{966585} uses a three-state Markov chain containing the following states:
\begin{itemize}
  \item \ac{los} (no shadowing)
  \item shadow (moderate shadowing conditions)
  \item deep shadow (heavy shadowing conditions)
\end{itemize}
In each state, the channel is modeled using the Loo distribution~\cite{1623307}. 
The overall received signal is described as a sum of the log-normally distributed direct component and the Rayleigh-distributed multipath component with parameters $\alpha$, $\psi$, and \MP.
The log-normal distribution is characterized by its mean $\alpha$ and standard deviation $\psi$, and the Rayleigh distribution by its average power \MP.
The specific values for $\alpha$, $\psi$, and \MP are chosen depending on the channel state and the satellite's elevation angle.
To train our joint encoder-decoder model, these values can be obtained from statistical experiments \cite{channel-params}.

The \ac{pdf} of the amplitude variations is as follows:
\begin{align}
  p(r) &= \frac{r}{b_0\sqrt{2\pi d_0}} \nonumber
  \int_{0}^{\infty}\frac{1}{s} \exp\left[-\frac{(\ln s - \mu)^2}{2d_0} - \frac{r^2 + s^2}{2b_0}\right] \\[1ex]
  &\hphantom{=\int_{0}^{\infty}} \cdot I_0\Big(\frac{rz}{b_0}\Big)\ ds,
\end{align}
where $s$ is the direct signal, and $I_0(\cdot)$ is the modified Bessel function of order zero that models electromagnetic wave propagation.
Further, $\mu, b_0,$ and $d_0$ are variables that can be derived from $\alpha$, $\psi$ and \MP as follows:
\begin{align}
\alpha &= 20\log_{10}(e^\mu) \\
  \psi &= 20\log_{10}(e^{\sqrt{d_0}}) \\
   \MP &= 10\log_{10}(2b_0)
\end{align}

We simulate the Loo distribution as described in \cite{DBLP:journals/ijscn/Perez-FontanMMPMMR08} to obtain $h$ based on the input values $\alpha$, $psi$, and $MP$.
That is, to compute the multipath vector, we generate two series of normally distributed variables and multiply them with \MP.
To compute the direct signal vector, we generate a series of normally distributed variables with mean $\alpha$ and standard deviation $\psi$, which we then convert to linear units and combine it with the multipath component using complex addition.
Note that, in contrast to \cite{DBLP:journals/ijscn/Perez-FontanMMPMMR08}, we do not model Doppler spread since we assume that the receiver is stationary and we do not model Doppler shift since the variations caused by Doppler shift are typically slow and can be corrected at the receiver~\cite{DBLP:journals/ijscn/Perez-FontanMMPMMR08}. 

Next, we consider the Gaussian noise $n$. We follow the approach presented in~\cite{satjscc}. 
We first compute the typical \ac{snr} values. 
To this end, we determine the distance $d$ between the satellite and the ground station based on the current elevation angle $\epsilon_0$ and Earth radius $R_E = 6378\,\textrm{km}$ as described in \cite{7506756}:
\begin{equation}
  d = R_E\Biggl(\sqrt{\Bigl(\frac{h + R_E}{R_E}\Bigr)^2 - \cos^{2}\epsilon_{0}} - \sin\epsilon_{0}\Biggr),
\end{equation}

Next, we compute thermal noise $N$ as follows:
\begin{equation}
  N=k \cdot T \cdot B,
\end{equation}
where $B$ and $T$ denote the bandwidth and noise temperature respectively and $k = 1.380649 \cdot 10^{-23}$ is the Boltzmann's constant.
We compute the noise temperature as the sum of the antenna temperature $T_a = 290 K$ and the receiver noise temperature $T_e$, which is determined as:
\begin{equation}
  T_e = T_0 \big(F_{\mathrm{sys}} - 1 \big),
\end{equation}
$T_0 = 290$\,K is the reference temperature and $F_{\textrm{sys}} = 2$\,dB denotes the receiver's noise figure.

Next, we compute the path loss $L$ using the Friis formula and determine the SNR depending on $L$ and $N$ as follows:
\begin{equation}
  \mathrm{SNR} = P_t + G_t + G_r - L - N,
\end{equation}
where $P_t$, $G_t$ and $G_r$ are input parameters denoting the transmitted power, the transmitter and the receiver gains respectively. 
The computed $\mathrm{SNR}$ value is then used to compute the noise power $\sigma^2$:
\begin{equation}
  \sigma^2 = \frac{P_{\mathrm{sig}}}{2 \cdot 10^{\frac{\mathrm{SNR}}{10}}}
\end{equation}
where $P_{\mathrm{sig}}$ is the normalized signal power.
Finally, we compute the noise vector $n$ as follows:
\begin{equation}
  n = \sigma \times \bigl[\mathcal{N}(0,1) + j * \mathcal{N}(0,1) \bigr] 
\end{equation}

\section{Evaluation}
\label{sec:evaluation}

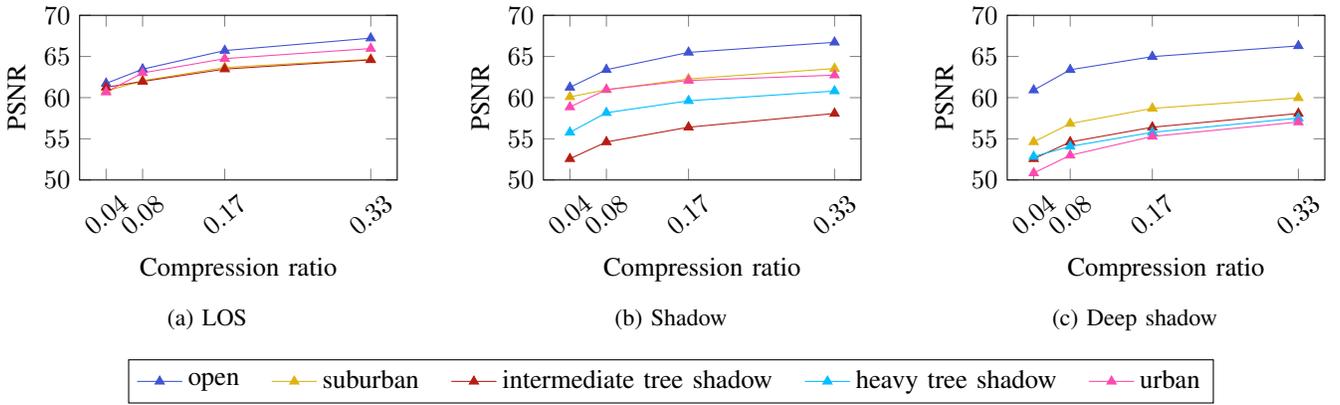
\begin{figure*}
  \begin{subfigure}{.32\linewidth}
  \centering
  \begin{tikzpicture}
    \begin{axis}[
        width=\linewidth,
        height=.65\linewidth,
        xlabel = {Compression ratio},
        ylabel = {PSNR},
        ymin = 50,
        ymax = 70,
        ylabel near ticks,
        xtick = data,
        x tick label style={
            rotate=40,
            /pgf/number format/fixed,
            /pgf/number format/precision=2,
        },
        x label style={below=0},
        ]

        \addplot[mark=triangle*,plt1] 
        plot 
        table[x expr=\thisrowno{0}, y expr=\thisrowno{1}, col sep=space] {plots/no_AF_40_open_los_los};

        \addplot[mark=triangle*,plt2] 
        plot 
        table[x expr=\thisrowno{0}, y expr=\thisrowno{1}, col sep=space] {plots/no_AF_40_suburban_los_los};

        \addplot[mark=triangle*,plt3] 
        plot 
        table[x expr=\thisrowno{0}, y expr=\thisrowno{1}, col sep=space] {plots/no_AF_40_intermediate_tree_shadow_los_los};
        
        
        \addplot[mark=triangle*,plt5] 
        plot 
        table[x expr=\thisrowno{0}, y expr=\thisrowno{1}, col sep=space] {plots/no_AF_40_urban_los_los};


    \end{axis}
\end{tikzpicture}
  \caption{LOS}
  \label{subfig:baseline_los}
\end{subfigure}
\hfill
\begin{subfigure}{.32\linewidth}
  \centering
  \begin{tikzpicture}
    \begin{axis}[
        width=\linewidth,
        height=.65\linewidth,
        xlabel = {Compression ratio},
        ylabel = {PSNR},
        ymin = 50,
        ymax = 70,
        ylabel near ticks,
        xtick = data,
        x tick label style={
            rotate=40,
            /pgf/number format/fixed,
            /pgf/number format/precision=2,
        },
        x label style={below=0},
        legend cell align={left},
        legend columns=5,
		legend style={/tikz/every even column/.append style={column sep=1em}},
        legend to name={legendenv},
        ]

        \addplot[mark=triangle*,plt1] 
        plot 
        table[x expr=\thisrowno{0}, y expr=\thisrowno{1}, col sep=space] {plots/no_AF_40_open_shadow_shadow};

        \addplot[mark=triangle*,plt2] 
        plot 
        table[x expr=\thisrowno{0}, y expr=\thisrowno{1}, col sep=space] {plots/no_AF_40_suburban_shadow_shadow};

        \addplot[mark=triangle*,plt3] 
        plot 
        table[x expr=\thisrowno{0}, y expr=\thisrowno{1}, col sep=space] {plots/no_AF_40_intermediate_tree_shadow_deep_shadow_deep_shadow};
        
        \addplot[mark=triangle*,plt4] 
        plot 
        table[x expr=\thisrowno{0}, y expr=\thisrowno{1}, col sep=space] {plots/no_AF_40_heavy_tree_shadow_shadow_shadow};
        
        \addplot[mark=triangle*,plt5] 
        plot 
        table[x expr=\thisrowno{0}, y expr=\thisrowno{1}, col sep=space] {plots/no_AF_40_urban_shadow_shadow};

        \legend{
            open,
            suburban, 
            intermediate tree shadow,
            heavy tree shadow,
            urban
            }

    \end{axis}
\end{tikzpicture}
  \caption{Shadow}
  \label{subfig:baseline_shadow}
\end{subfigure}
\hfill
\begin{subfigure}{.32\linewidth}
  \centering
  \begin{tikzpicture}
    \begin{axis}[
        width=\linewidth,
        height=.65\linewidth,
        xlabel = {Compression ratio},
        ylabel = {PSNR},
        ymin = 50,
        ymax = 70,
        ylabel near ticks,
        xtick = data,
        x tick label style={
            rotate=40,
            /pgf/number format/fixed,
            /pgf/number format/precision=2,
        },
        x label style={below=0},
        ]

        \addplot[mark=triangle*,plt1] 
        plot 
        table[x expr=\thisrowno{0}, y expr=\thisrowno{1}, col sep=space] {plots/no_AF_40_open_deep_shadow_deep_shadow};

        \addplot[mark=triangle*,plt2] 
        plot 
        table[x expr=\thisrowno{0}, y expr=\thisrowno{1}, col sep=space] {plots/no_AF_40_suburban_deep_shadow_deep_shadow};

        \addplot[mark=triangle*,plt3] 
        plot 
        table[x expr=\thisrowno{0}, y expr=\thisrowno{1}, col sep=space] {plots/no_AF_40_intermediate_tree_shadow_deep_shadow_deep_shadow};
        
        \addplot[mark=triangle*,plt4] 
        plot 
        table[x expr=\thisrowno{0}, y expr=\thisrowno{1}, col sep=space] {plots/no_AF_40_heavy_tree_shadow_deep_shadow_deep_shadow};
        
        \addplot[mark=triangle*,plt5] 
        plot 
        table[x expr=\thisrowno{0}, y expr=\thisrowno{1}, col sep=space] {plots/no_AF_40_urban_deep_shadow_deep_shadow};


    \end{axis}
\end{tikzpicture}
  \caption{Deep shadow}
  \label{subfig:baseline_deep_shadow}
\end{subfigure}

\vspace{1em}
\centering
\ref{legendenv}

\caption{\ac{psnr} achieved by the \baseline for different environments, states, and compression ratios with 40\textdegree{} elevation angle.}
\label{fig:different_scenes_40}
\end{figure*}

\begin{table}[b!]
  \caption{Channel Parameters}
  \label{tab:channel_parameters}

  \centering
	\begin{tabular}{ll}
		\toprule
    Parameter & Value \\
    \midrule
		Orbit height & 150\,km \\
		Carrier frequency & 2150\,MHz \\
    Transmitted power & 1\,W \\
    Satellite antenna gain & 6\,dBi \\
    Ground station antenna gain & 35\,dBi \\
		Receive channel bandwidth & 750\,kHz \\
		Noise figure & 2\,dB \\
		\bottomrule
	\end{tabular}
\end{table}

In this section, we evaluate our mechanism (hereafter: \ours) and compare the results against a mechanism that also uses \ac{jscc} but employs a network architecture without attention modules, using a separate network for each channel condition (hereafter: \baseline).

Both mechanisms are trained using \sentinelii multi-spectral images (cf. \Cref{sec:system_model}) from the region of Serbia in summer, which were extracted from the BigEarth dataset~\cite{sumbul2019bigearthnet,Sumbul2021}.
After filtering out cloudy images, which do not contribute a valuable signal, 14,439 images remain, which we further split into training, validation, and test sets.
We use cubic interpolation to extrapolate all spectral bands to the same image resolution and represent pixel values as normalized values between 0 and 1.

During training, we set the batch size to 32; the learning rate is $10^{-3}$ and adjusted to $10^{-4}$ after 500 epochs; and Adam is used as a stochastic gradient descent.
All code was written using Keras \cite{keras} and Tensorflow \cite{tensorflow}.

To represent different channel conditions, we use statistical measurements for $\alpha$, $\psi$, and \MP obtained in different environments and channel states \cite{channel-params}.
These measurements were conducted for a range of elevation angles from 40\textdegree to 80\textdegree.
We further calculate the expected \ac{snr} as described in \Cref{sub:channel_model}.
As environments, we consider \emph{open, suburban, intermediate tree shadow, heavy tree shadow,} and \emph{urban.}
The channel states are differentiated with respect to their shadowing conditions as \emph{\acf{los}, shadow,} and \emph{deep shadow.}
In addition, we consider a number of compression ratios $k/n$ ranging from $0.04$ to $0.33$, as the amount of tolerable compression differs depending on the use case.

\begin{figure*}
  \begin{subfigure}{.32\linewidth}
  \centering
  \begin{tikzpicture}
    \begin{axis}[
      ybar,
      width=\linewidth,
      height=.65\linewidth,
      ylabel near ticks,
      xlabel near ticks,
      xmajorticks=false,
      every axis plot/.append style={
                bar width=8pt,
                bar shift=0pt,
                fill
              },
      ymajorgrids=true,
      ylabel={PSNR},
      ymin = 50,
        ymax = 70,
      enlargelimits=.2,
      xlabel={Environments},
      symbolic x coords ={%
          open,
          suburban,
          intermediate tree-shadow,
          urban,
      },
      legend cell align={left},
      legend style={column sep=5pt},
      legend to name={legendangles},
      legend columns=2,
      xticklabel=\empty,
      ]
      
      \legend{
        40\textdegree{},
        80\textdegree{}
      }

      \addplot[
        bar shift=-5pt,fill=white,postaction={
        pattern=north east lines, pattern color=black
    }] coordinates { (open, 0) };
      \addplot[bar shift=5pt,fill=white] coordinates { (open, 0) };

      \addplot[
        bar shift=-5pt,fill=plt1!70,postaction={
        pattern=north east lines, pattern color=black
    }] coordinates { (open, 61.73067188214748) };
      \addplot[bar shift=5pt,fill=plt1!70] coordinates { (open, 62.46548778477633) };

      \addplot[
      bar shift=-5pt,fill=plt2!70,postaction={
        pattern=north east lines, pattern color=black
    }] coordinates { (suburban, 60.805258435946925) };
      \addplot[bar shift=5pt,fill=plt2!70] coordinates { (suburban, 60.92828818309016) };
      
      \addplot[
        bar shift=-5pt,fill=plt3!70,postaction={
        pattern=north east lines, pattern color=black
    }] coordinates { (intermediate tree-shadow, 61.2610153054711) };
      \addplot[bar shift=5pt,fill=plt3!70] coordinates { (intermediate tree-shadow, 60.29798046606024) };
      
      \addplot[
        bar shift=-5pt,fill=plt5!70,postaction={
        pattern=north east lines, pattern color=black
    }] coordinates { (urban, 60.67812370860495) };
      \addplot[bar shift=5pt,fill=plt5!70] coordinates { (urban, 61.670784487418885) };
      
    \end{axis} 
  \end{tikzpicture}
  \caption{LOS, compression ratio 0.04}
  \label{subfig:base4080_los_04}
\end{subfigure}
\hfill
\begin{subfigure}{.32\linewidth}
  \centering
  \begin{tikzpicture}
    \begin{axis}[
      ybar,
      width=\linewidth,
      height=.65\linewidth,
      ylabel near ticks,
      xlabel near ticks,
      xmajorticks=false,
      every axis plot/.append style={
                bar width=8pt,
                bar shift=0pt,
                fill
              },
      ymajorgrids=true,
      ylabel={PSNR},
      ymin = 50,
        ymax = 70,
      enlargelimits=.2,
      xlabel={Environments},
      symbolic x coords ={%
          open,
          suburban,
          intermediate tree-shadow,
          heavy tree-shadow,
          urban,
      },
      xticklabel=\empty,
      ]
  
      \addplot[
        bar shift=-5pt,fill=white,postaction={
        pattern=north east lines, pattern color=black
    }] coordinates { (open, 0) };
      \addplot[bar shift=5pt,fill=white] coordinates { (open, 0) };

      \addplot[
        bar shift=-5pt,fill=plt1!70,postaction={
        pattern=north east lines, pattern color=black
    }] coordinates { (open, 61.242307462063216) };
      \addplot[bar shift=5pt,fill=plt1!70] coordinates { (open, 62.43216132742652) };

      \addplot[
      bar shift=-5pt,fill=plt2!70,postaction={
        pattern=north east lines, pattern color=black
    }] coordinates { (suburban, 60.0867862998761) };
      \addplot[bar shift=5pt,fill=plt2!70] coordinates { (suburban, 61.12110295624692) };
      
      \addplot[
        bar shift=-5pt,fill=plt3!70,postaction={
        pattern=north east lines, pattern color=black
    }] coordinates { (intermediate tree-shadow, 57.62850182792831) };
      \addplot[bar shift=5pt,fill=plt3!70] coordinates { (intermediate tree-shadow, 59.88167207543309) };
      
      \addplot[
        bar shift=-5pt,fill=plt4!70,postaction={
        pattern=north east lines, pattern color=black
    }] coordinates { (heavy tree-shadow, 55.79653619027105) };
      \addplot[bar shift=5pt,fill=plt4!70] coordinates { (heavy tree-shadow, 58.7045005554619) };
      
      \addplot[
        bar shift=-5pt,fill=plt5!70,postaction={
        pattern=north east lines, pattern color=black
    }] coordinates { (urban, 58.862160183144205) };
      \addplot[bar shift=5pt,fill=plt5!70] coordinates { (urban, 59.03699565522086) };
      
    \end{axis} 
  \end{tikzpicture}
  \caption{Shadow, compression ratio 0.04}
  \label{subfig:base4080_shadow_04}
\end{subfigure}
\hfill
\begin{subfigure}{.32\linewidth}
  \centering
  \begin{tikzpicture}
    \begin{axis}[
      ybar,
      width=\linewidth,
      height=.65\linewidth,
      ylabel near ticks,
      xlabel near ticks,
      xmajorticks=false,
      every axis plot/.append style={
                bar width=8pt,
                bar shift=0pt,
                fill
              },
      ymajorgrids=true,
      ylabel={PSNR},
      ymin = 50,
        ymax = 70,
      enlargelimits=.2,
      xlabel={Environments},
      symbolic x coords ={%
          open,
          suburban,
          intermediate tree-shadow,
          heavy tree-shadow,
          urban,
      },
      xticklabel=\empty,
      ]
  
      \addplot[
        bar shift=-5pt,fill=white,postaction={
        pattern=north east lines, pattern color=black
    }] coordinates { (open, 0) };
      \addplot[bar shift=5pt,fill=white] coordinates { (open, 0) };

      \addplot[
        bar shift=-5pt,fill=plt1!70,postaction={
        pattern=north east lines, pattern color=black
    }] coordinates { (open, 60.915204772562106) };
      \addplot[bar shift=5pt,fill=plt1!70] coordinates { (open, 60.941951355437276) };

      \addplot[
      bar shift=-5pt,fill=plt2!70,postaction={
        pattern=north east lines, pattern color=black
    }] coordinates { (suburban, 54.626502435886835) };
      \addplot[bar shift=5pt,fill=tol2!70] coordinates { (suburban, 61.017142451016966) };
      
      \addplot[
        bar shift=-5pt,fill=plt3!70,postaction={
        pattern=north east lines, pattern color=black
    }] coordinates { (intermediate tree-shadow, 52.57299641191118) };
      \addplot[bar shift=5pt,fill=plt3!70] coordinates { (intermediate tree-shadow, 60.14165345147827) };
      
      \addplot[
        bar shift=-5pt,fill=plt4!70,postaction={
        pattern=north east lines, pattern color=black
    }] coordinates { (heavy tree-shadow, 52.84660367941581) };
      \addplot[bar shift=5pt,fill=plt4!70] coordinates { (heavy tree-shadow, 55.73117874454389) };
      
      \addplot[
        bar shift=-5pt,fill=plt5!70,postaction={
        pattern=north east lines, pattern color=black
    }] coordinates { (urban, 50.84575762551614) };
      \addplot[bar shift=5pt,fill=plt5!70] coordinates { (urban, 57.86157808541486) };
      
    \end{axis} 
  \end{tikzpicture}
  \caption{Deep shadow, compression ratio 0.04}
  \label{subfig:base4080_deep_shadow_04}
\end{subfigure}

\vspace{1em}

\begin{subfigure}{.32\linewidth}
  \centering
  \begin{tikzpicture}
    \begin{axis}[
      ybar,
      width=\linewidth,
      height=.65\linewidth,
      ylabel near ticks,
      xlabel near ticks,
      xmajorticks=false,
      every axis plot/.append style={
                bar width=8pt,
                bar shift=0pt,
                fill
              },
      ymajorgrids=true,
      ylabel={PSNR},
      ymin = 50,
        ymax = 70,
      enlargelimits=.2,
      xlabel={Environments},
      symbolic x coords ={%
          open,
          suburban,
          intermediate tree-shadow,
          urban,
      },
      xticklabel=\empty,
      ]

      \addplot[
        bar shift=-5pt,fill=white,postaction={
        pattern=north east lines, pattern color=black
    }] coordinates { (open, 0) };
      \addplot[bar shift=5pt,fill=white] coordinates { (open, 0) };

      \addplot[
        bar shift=-5pt,fill=plt1!70,postaction={
        pattern=north east lines, pattern color=black
    }] coordinates { (open, 67.22762412790338) };
      \addplot[bar shift=5pt,fill=plt1!70] coordinates { (open, 67.94286665070874) };

      \addplot[
      bar shift=-5pt,fill=plt2!70,postaction={
        pattern=north east lines, pattern color=black
    }] coordinates { (suburban, 64.65365825355423) };
      \addplot[bar shift=5pt,fill=plt2!70] coordinates { (suburban, 65.65526368893802) };
      
      \addplot[
        bar shift=-5pt,fill=plt3!70,postaction={
        pattern=north east lines, pattern color=black
    }] coordinates { (intermediate tree-shadow, 64.59177795781159) };
      \addplot[bar shift=5pt,fill=plt3!70] coordinates { (intermediate tree-shadow, 63.51533179513708) };
      
      \addplot[
        bar shift=-5pt,fill=plt5!70,postaction={
        pattern=north east lines, pattern color=black
    }] coordinates { (urban, 65.96625190247083) };
      \addplot[bar shift=5pt,fill=plt5!70] coordinates { (urban, 66.64719138769036) };
      
    \end{axis} 
  \end{tikzpicture}
  \caption{LOS, compression ratio 0.33}
  \label{subfig:base4080_los_33}
\end{subfigure}
\hfill
\begin{subfigure}{.32\linewidth}
  \centering
  \begin{tikzpicture}
    \begin{axis}[
      ybar,
      width=\linewidth,
      height=.65\linewidth,
      ylabel near ticks,
      xlabel near ticks,
      xmajorticks=false,
      every axis plot/.append style={
                bar width=8pt,
                bar shift=0pt,
                fill
              },
      ymajorgrids=true,
      ylabel={PSNR},
      ymin = 50,
        ymax = 70,
      enlargelimits=.2,
      xlabel={Environments},
      symbolic x coords ={%
          open,
          suburban,
          intermediate tree-shadow,
          heavy tree-shadow,
          urban,
      },
      xticklabel=\empty,
      ]

      \addplot[
        bar shift=-5pt,fill=white,postaction={
        pattern=north east lines, pattern color=black
    }] coordinates { (open, 0) };
      \addplot[bar shift=5pt,fill=white] coordinates { (open, 0) };

      \addplot[
        bar shift=-5pt,fill=plt1!70,postaction={
        pattern=north east lines, pattern color=black
    }] coordinates { (open, 66.72378601897488) };
      \addplot[bar shift=5pt,fill=plt1!70] coordinates { (open, 68.05071847613172) };

      \addplot[
      bar shift=-5pt,fill=plt2!70,postaction={
        pattern=north east lines, pattern color=black
    }] coordinates { (suburban, 63.5362946198049) };
      \addplot[bar shift=5pt,fill=plt2!70] coordinates { (suburban, 65.61713528754578) };
      
      \addplot[
        bar shift=-5pt,fill=plt3!70,postaction={
        pattern=north east lines, pattern color=black
    }] coordinates { (intermediate tree-shadow, 61.85098102603866) };
      \addplot[bar shift=5pt,fill=plt3!70] coordinates { (intermediate tree-shadow, 63.182928753310684) };
      
      \addplot[
        bar shift=-5pt,fill=plt4!70,postaction={
        pattern=north east lines, pattern color=black
    }] coordinates { (heavy tree-shadow, 60.80235241808859) };
      \addplot[bar shift=5pt,fill=plt4!70] coordinates { (heavy tree-shadow, 62.51972868765633) };
      
      \addplot[
        bar shift=-5pt,fill=plt5!70,postaction={
        pattern=north east lines, pattern color=black
    }] coordinates { (urban, 62.73554608081638) };
      \addplot[bar shift=5pt,fill=plt5!70] coordinates { (urban, 62.9654737199328) };
      
    \end{axis} 
  \end{tikzpicture}
  \caption{Shadow, compression ratio 0.33}
  \label{subfig:base4080_shadow_33}
\end{subfigure}
\hfill
\begin{subfigure}{.32\linewidth}
  \centering
  \begin{tikzpicture}
    \begin{axis}[
      ybar,
      width=\linewidth,
      height=.65\linewidth,
      ylabel near ticks,
      xlabel near ticks,
      xmajorticks=false,
      every axis plot/.append style={
                bar width=8pt,
                bar shift=0pt,
                fill
              },
      ymajorgrids=true,
      ylabel={PSNR},
      ymin = 50,
        ymax = 70,
      enlargelimits=.2,
      xlabel={Environments},
      symbolic x coords ={%
          open,
          suburban,
          intermediate tree-shadow,
          heavy tree-shadow,
          urban,
      },
      xticklabel=\empty,
      ]

      \addplot[
        bar shift=-5pt,fill=white,postaction={
        pattern=north east lines, pattern color=black
    }] coordinates { (open, 0) };
      \addplot[bar shift=5pt,fill=white] coordinates { (open, 0) };

      \addplot[
        bar shift=-5pt,fill=plt1!70,postaction={
        pattern=north east lines, pattern color=black
    }] coordinates { (open, 66.28719931983984) };
      \addplot[bar shift=5pt,fill=plt1!70] coordinates { (open, 67.76665563361574) };

      \addplot[
      bar shift=-5pt,fill=plt2!70,postaction={
        pattern=north east lines, pattern color=black
    }] coordinates { (suburban, 59.97375970276803) };
      \addplot[bar shift=5pt,fill=plt2!70] coordinates { (suburban, 64.41231830717751) };
      
      \addplot[
        bar shift=-5pt,fill=plt3!70,postaction={
        pattern=north east lines, pattern color=black
    }] coordinates { (intermediate tree-shadow, 58.080011031063236) };
      \addplot[bar shift=5pt,fill=plt3!70] coordinates { (intermediate tree-shadow, 63.39092156973394) };
      
      \addplot[
        bar shift=-5pt,fill=plt4!70,postaction={
        pattern=north east lines, pattern color=black
    }] coordinates { (heavy tree-shadow, 57.512479690636944) };
      \addplot[bar shift=5pt,fill=plt4!70] coordinates { (heavy tree-shadow, 60.45771389019385) };
      
      \addplot[
        bar shift=-5pt,fill=plt5!70,postaction={
        pattern=north east lines, pattern color=black
    }] coordinates { (urban, 57.03474187203291) };
      \addplot[bar shift=5pt,fill=plt5!70] coordinates { (urban, 62.0518730036318) };
      
    \end{axis} 
  \end{tikzpicture}
  \caption{Deep shadow, compression ratio 0.33}
  \label{subfig:base4080_deep_shadow_33}
\end{subfigure}

\vspace{1em}
\centering
\ref{legendangles}
\tikz{%
  \node[draw=black, inner sep=3.55pt] {
    \tikz{\node[fill=plt1] {};} open \hspace{1em}
    \tikz{\node[fill=plt2] {};} suburban \hspace{1em}
    \tikz{\node[fill=plt3] {};} intermediate tree shadow \hspace{1em}
    \tikz{\node[fill=plt4] {};} heavy tree shadow \hspace{1em}
    \tikz{\node[fill=plt5] {};} urban
  };
}

\caption{\ac{psnr} achieved by the \baseline for different environments and 40\textdegree{} vs. 80\textdegree{} elevation angle.}
\label{fig:elevation_angles}
\end{figure*}
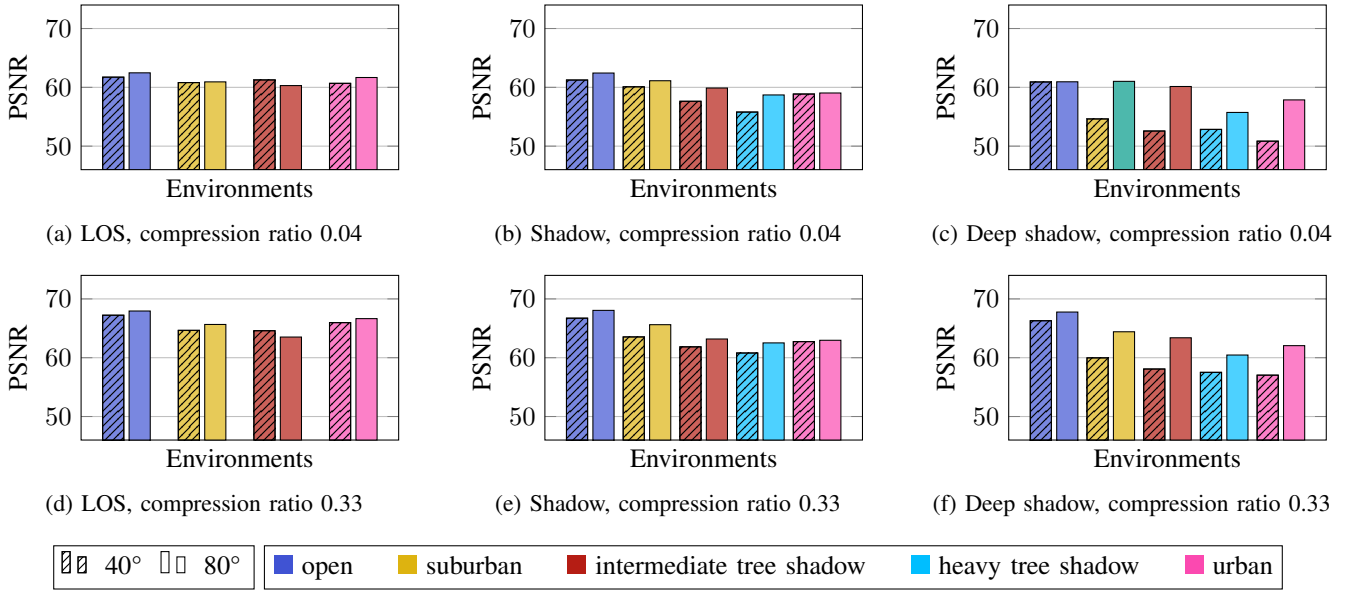

We consider three evaluation scenarios:
\begin{enumerate}
  \item We evaluate how the \baseline performs in different environments, channel states, compression ratios, and elevation angles.
  \item We compare \ours against the \baseline to evaluate the impact of attention modules on the performance.
  \item We compare \ours against the \baseline in a scenario where the detected channel state differs from the actual channel state, and thus, the wrong network or attention module parametrization are used.
\end{enumerate}

For all scenarios, we compare the image quality measured using \ac{psnr} -- higher values indicate better image quality.

While adding attention modules incurs additional overhead, it is negatable in our case. 
The attention module parameters comprise only $0.25\%$ of all model parameters in \ours.  
However, a single \ours model captures a range of channel conditions, significantly reducing the overall storage overhead.
Specifically, we train a single \ours model for the three channel states and different \ac{snr} values.
Different models are still used for different environments and compression ratios, as we assume these parameters to change less frequently.
Due to the use of attention modules, we expect \ours to perform slightly worse in terms of \ac{psnr} when compared to a custom tailored model.
The main goal of our evaluation is, therefore, to determine whether the loss in \ac{psnr} is negligible when compared to the gain in storage efficiency.

First, we discuss the \baseline's results for different environments, states, and compression ratios $k/n$ with 40\textdegree{} elevation angle shown in \Cref{fig:different_scenes_40}. 
The $x$-axis shows different compression ratios and the $y$-axis the achieved \ac{psnr} values.
It can be seen that the results strongly depend on the environment.
As expected, the best \ac{psnr} values are achieved in the open environment and the worst in more challenging environments, such as urban or intermediate tree shadow.
It also can be seen that the results significantly depend on shadowing conditions. 
While for deep shadow (\Cref{subfig:baseline_deep_shadow}), the performance varies considerably between different environments, 
the differences become less pronounced as shadowing conditions improve (\Cref{subfig:baseline_shadow,subfig:baseline_los}).

Next, we consider the differences between 40\textdegree{} and 80\textdegree{} elevation angles for the \baseline. 
The results for compression ratios of $k/n = 0.04$ and $k/n = 0.33$ are presented in \Cref{fig:elevation_angles}.
Again, the results depend both on the environment and the shadowing state.
While in \ac{los} conditions (\Cref{subfig:base4080_los_04,subfig:base4080_los_33}), the differences between 40\textdegree{} and 80\textdegree{} elevation angle remain negligible across different environments, 
stronger performance variations can be seen for the more challenging states shadow (\Cref{subfig:base4080_shadow_04,subfig:base4080_shadow_33}) and deep shadow (\Cref{subfig:base4080_deep_shadow_04,subfig:base4080_deep_shadow_33}).
In addition, we observe that the chosen compression ratio influences the results depending on the elevation angle.
More specifically, the results suggest that moderate compression ratios should be chosen in case of low elevation angles.

\begin{figure*}
  \begin{subfigure}{.48\linewidth}
  \centering
  \begin{tikzpicture}
    \begin{axis}[
        width=\linewidth,
        height=.56\linewidth,
        xlabel = {Compression ratio},
        ylabel = {PSNR},
        legend cell align={left},
        ymin = 50,
        ymax = 70,
		legend columns=3,
		legend style={/tikz/every even column/.append style={column sep=1em}},
        ylabel near ticks,
        xtick = data,
        x tick label style={
            rotate=0,
            /pgf/number format/fixed,
            /pgf/number format/precision=2,
        },
        x label style={below=0},
        legend to name={legendurban}
        ]

        \addplot[mark=triangle*,plt1, thick] 
        plot 
        table[x expr=\thisrowno{0}, y expr=\thisrowno{1}, col sep=space] {plots//no_AF_40_urban_los_los};
        
        \addplot[mark=triangle*,plt2, thick] 
        plot 
        table[x expr=\thisrowno{0}, y expr=\thisrowno{1}, col sep=space] {plots//no_AF_40_urban_shadow_shadow};

        \addplot[mark=triangle*,plt5, thick] 
        plot 
        table[x expr=\thisrowno{0}, y expr=\thisrowno{1}, col sep=space] {plots//no_AF_40_urban_deep_shadow_deep_shadow};

        \addplot[mark=square*,plt1, thick, dotted] 
        plot 
        table[x expr=\thisrowno{0}, y expr=\thisrowno{1}, col sep=space] {plots//AF_SNR_40_urban_los_los};
        
        \addplot[mark=square*,plt2, thick, dotted] 
        plot 
        table[x expr=\thisrowno{0}, y expr=\thisrowno{1}, col sep=space] {plots//AF_SNR_40_urban_shadow_shadow};

        \addplot[mark=square*,plt5, thick, dotted] 
        plot 
        table[x expr=\thisrowno{0}, y expr=\thisrowno{1}, col sep=space] {plots//AF_SNR_40_urban_deep_shadow_deep_shadow};

        \legend{
            \baseline{} / LOS,
            \baseline{} / shadow, 
            \baseline{} / deep shadow,
            \ours{} / LOS,
            \ours{} / shadow,
            \ours{} / deep shadow
            }

    \end{axis}
\end{tikzpicture}
  \caption{40\textdegree{} elevation angle}
\end{subfigure}
\hfill
\begin{subfigure}{.48\linewidth}
  \centering
  \begin{tikzpicture}
    \begin{axis}[
        width=\linewidth,
        height=.56\linewidth,
        xlabel = {Compression ratio},
        ylabel = {PSNR},
        ymin = 50,
        ymax = 70,
        ylabel near ticks,
        xtick = data,
        x tick label style={
            rotate=0,
            /pgf/number format/fixed,
            /pgf/number format/precision=2,
        },
        x label style={below=0},
        ]

        \addplot[mark=triangle*,plt1, thick] 
        plot 
        table[x expr=\thisrowno{0}, y expr=\thisrowno{1}, col sep=space] {plots//no_AF_80_urban_los_los};
        
        \addplot[mark=triangle*,plt2, thick] 
        plot 
        table[x expr=\thisrowno{0}, y expr=\thisrowno{1}, col sep=space] {plots//no_AF_80_urban_shadow_shadow};

        \addplot[mark=triangle*,plt5, thick] 
        plot 
        table[x expr=\thisrowno{0}, y expr=\thisrowno{1}, col sep=space] {plots//no_AF_80_urban_deep_shadow_deep_shadow};

        \addplot[mark=square*,plt1, thick, dotted] 
        plot 
        table[x expr=\thisrowno{0}, y expr=\thisrowno{1}, col sep=space] {plots//AF_SNR_80_urban_los_los};
        
        \addplot[mark=square*,plt2, thick, dotted] 
        plot 
        table[x expr=\thisrowno{0}, y expr=\thisrowno{1}, col sep=space] {plots//AF_SNR_80_urban_shadow_shadow};

        \addplot[mark=square*,plt5, thick, dotted] 
        plot 
        table[x expr=\thisrowno{0}, y expr=\thisrowno{1}, col sep=space] {plots//AF_SNR_80_urban_deep_shadow_deep_shadow};


    \end{axis}
\end{tikzpicture}
  \caption{80\textdegree{} elevation angle}
\end{subfigure}

\vspace{1em}
\centering
\ref{legendurban}

\caption{\ac{psnr} of \ours vs. \baseline in an urban environment.}
\label{fig:af_vs_no_af}
\end{figure*}

\begin{figure*}[t!]
  \begin{subfigure}{.48\linewidth}
  \centering
  \begin{tikzpicture}
    \begin{axis}[
        width=\linewidth,
        height=.56\linewidth,
        xlabel = {Compression ratio},
        ylabel = {PSNR},
        legend cell align={left},
        ymin = 25,
        ymax = 75,
        ytick distance = 10,
		legend columns=4,
		legend style={/tikz/every even column/.append style={column sep=1em}},
        ylabel near ticks,
        xtick = data,
        x tick label style={
            rotate=0,
            /pgf/number format/fixed,
            /pgf/number format/precision=2,
        },
        x label style={below=0},
        legend to name={legenderror}
        ]

        \addplot[mark=triangle*,plt1, thick] 
        plot 
        table[x expr=\thisrowno{0}, y expr=\thisrowno{1}, col sep=space] {plots/no_AF_40_urban_los_los};
        
        \addplot[mark=triangle*,mark options=solid,plt1, thick, dashed] 
        plot 
        table[x expr=\thisrowno{0}, y expr=\thisrowno{1}, col sep=space] {plots/no_AF_40_40_urban_deep_shadow_los};

        \addplot[mark=triangle*,plt2, thick] 
        plot 
        table[x expr=\thisrowno{0}, y expr=\thisrowno{1}, col sep=space] {plots/no_AF_80_urban_los_los};

        \addplot[mark=triangle*,mark options=solid,plt2, thick, dashed] 
        plot 
        table[x expr=\thisrowno{0}, y expr=\thisrowno{1}, col sep=space] {plots/no_AF_80_80_urban_deep_shadow_los};

        \addplot[mark=square*,plt3, thick] 
        plot 
        table[x expr=\thisrowno{0}, y expr=\thisrowno{1}, col sep=space] {plots/AF_SNR_40_urban_los_los};

        \addplot[mark=square*,mark options=solid,plt3, thick, dashed] 
        plot 
        table[x expr=\thisrowno{0}, y expr=\thisrowno{1}, col sep=space] {plots/AF_SNR_40_40_urban_deep_shadow_los};

        \addplot[mark=square*,plt4, thick] 
        plot 
        table[x expr=\thisrowno{0}, y expr=\thisrowno{1}, col sep=space] {plots/AF_SNR_80_urban_los_los};

        \addplot[mark=square*,mark options=solid,plt4, thick, dashed] 
        plot 
        table[x expr=\thisrowno{0}, y expr=\thisrowno{1}, col sep=space] {plots/AF_SNR_80_80_urban_deep_shadow_los};


        \legend{
            \baseline{} / correct / 40\textdegree,
            \baseline{} / error / 40\textdegree, 
            \baseline{} / correct / 80\textdegree,
            \baseline{} / error / 80\textdegree,
            \ours{} / correct / 40\textdegree,
            \ours{} / error / 40\textdegree, 
            \ours{} / correct / 80\textdegree,
            \ours{} / error / 80\textdegree,
            }

    \end{axis}
\end{tikzpicture}
  \caption{LOS instead of deep shadow (better than expected)}
  \label{fig:different_state_same_snr_better}
\end{subfigure}
\hfill
\begin{subfigure}{.48\linewidth}
  \centering
  \begin{tikzpicture}
    \begin{axis}[
        width=\linewidth,
        height=.56\linewidth,
        xlabel = {Compression ratio},
        ylabel = {PSNR},
        ymin = 25,
        ymax = 75,
        ytick distance = 10,
        ylabel near ticks,
        xtick = data,
        x tick label style={
            rotate=0,
            /pgf/number format/fixed,
            /pgf/number format/precision=2,
        },
        x label style={below=0},
        ]

        \addplot[mark=triangle*,plt1, thick] 
        plot 
        table[x expr=\thisrowno{0}, y expr=\thisrowno{1}, col sep=space] {plots/no_AF_40_urban_deep_shadow_deep_shadow};
        
        \addplot[mark=triangle*,mark options=solid,plt1, thick, dashed] 
        plot 
        table[x expr=\thisrowno{0}, y expr=\thisrowno{1}, col sep=space] {plots/no_AF_40_40_urban_los_deep_shadow};

        \addplot[mark=triangle*,plt2, thick] 
        plot 
        table[x expr=\thisrowno{0}, y expr=\thisrowno{1}, col sep=space] {plots/no_AF_80_urban_deep_shadow_deep_shadow};

        \addplot[mark=triangle*,mark options=solid,plt2, thick, dashed] 
        plot 
        table[x expr=\thisrowno{0}, y expr=\thisrowno{1}, col sep=space] {plots/no_AF_80_80_urban_los_deep_shadow};

        \addplot[mark=square*,plt3, thick] 
        plot 
        table[x expr=\thisrowno{0}, y expr=\thisrowno{1}, col sep=space] {plots/AF_SNR_40_urban_deep_shadow_deep_shadow};

        \addplot[mark=square*,plt3,mark options=solid, thick, dashed] 
        plot 
        table[x expr=\thisrowno{0}, y expr=\thisrowno{1}, col sep=space] {plots/AF_SNR_40_40_urban_los_deep_shadow};

        \addplot[mark=square*,plt4, thick] 
        plot 
        table[x expr=\thisrowno{0}, y expr=\thisrowno{1}, col sep=space] {plots/AF_SNR_80_urban_deep_shadow_deep_shadow};

        \addplot[mark=square*,plt4,mark options=solid, thick, dashed] 
        plot 
        table[x expr=\thisrowno{0}, y expr=\thisrowno{1}, col sep=space] {plots/AF_SNR_80_80_urban_los_deep_shadow};


    \end{axis}
\end{tikzpicture}
  \caption{Deep shadow instead of LOS (worse than expected)}
  \label{fig:different_state_same_snr_worse}
\end{subfigure}

\vspace{1em}
\centering
\ref{legenderror}

\caption{Channel conditions differing from the expected state in an urban environment.}
\label{fig:different_state_same_snr}
\end{figure*}

Next, we evaluate how the use of attention modules influences the performance.
We compare the neural network architectures with attention modules (\ours) and without attention modules (\baseline).
For our evaluation, we choose an urban environment since it shows stronger variability for different states and elevation angles.
The results are presented in \Cref{fig:af_vs_no_af}.
We can see that the performance of \ours depends on the compression ratio and the channel state and that the elevation angle plays a less important role. 
When stronger compression is applied, the \ac{psnr} values are similar to that of the \baseline, which uses separate neural network models for different conditions. 
In case of \ac{los} with a 40\textdegree{} elevation angle, \ours exhibits even slightly better performance than the \baseline.
For less aggressive compression, \ours shows slightly inferior results.
The biggest performance gap is observed in deep shadow conditions and the smallest in shadow conditions. 
This gap becomes even smaller for shadow conditions when the elevation angle is set to 80\textdegree{}. 
However, there is no significant difference between 40\textdegree{} and 80\textdegree{} elevation for the other channel states.

Finally, we evaluate what happens when there is a mismatch between the estimated channel parameters and the real channel conditions. 
The underlying assumption is that both the satellite and the ground station constantly monitor the observed channel conditions and either select the most suitable neural network model (in case of the \baseline) or the most suitable attention module parameters (in case of \ours).
Therefore, it is possible that the wrong model is selected, which may lead to reduced performance.
The results are presented in \Cref{fig:different_state_same_snr}.
In \Cref{fig:different_state_same_snr_better}, assume the satellite and ground station both assume a deep shadow state but the actual state is \ac{los}. 
That is, the channel conditions are better than expected. 
We compare the results for the correct channel estimation (solid lines) and wrong channel estimations (dashed lines).
The results show that both \ours and the \baseline are not able to benefit from the better channel state and experience noticeable performance degradation.
The results differ a lot when comparing 40\textdegree{} and 80\textdegree{} elevation angles.
While for 80\textdegree{}, both architectures achieve similar results, 
\ours performs significantly better for 40\textdegree{} elevation angle.
In \Cref{fig:different_state_same_snr_worse}, we consider the opposite case: 
A \ac{los} state is expected and the actual channel state is deep shadow. 
Similar to the previous example, the performance degrades in the case of the wrong state estimation.
For both elevation angles, \ours outperforms the \baseline.
Therefore, we observe that the use of attention module can bring additional benefits besides storage efficiency in cases where wrong channel conditions are assumed.

\section{Conclusion}
\label{sec:conclusion}

Earth observation missions employing modern sensor technology can produce vast amounts of data, complicating the design of efficient communication mechanisms when met with the harsh channel conditions typical for space scenarios.
Deep \Acf{jscc} can be applied to tackle these challenges.

In this paper, we have proposed an advanced \ac{jscc} design that takes into account a realistic channel model for small satellite communication.
Furthermore, by augmenting an encoder-decoder neuronal network model architecture with attention modules, a single model can be parametrized for different channel conditions, significantly reducing complexity when compared to separate models per per channel condition.

Our evaluation results show that the attention-module-enhanced model performs similar to individual networks.
In particular, our approach even outperforms individual models in cases where there is a mismatch between the detected and actual channel conditions.

\bibliographystyle{IEEEtran}
\bibliography{references}

\begin{thebibliography}{10}
\providecommand{\url}[1]{#1}
\csname url@samestyle\endcsname
\providecommand{\newblock}{\relax}
\providecommand{\bibinfo}[2]{#2}
\providecommand{\BIBentrySTDinterwordspacing}{\spaceskip=0pt\relax}
\providecommand{\BIBentryALTinterwordstretchfactor}{4}
\providecommand{\BIBentryALTinterwordspacing}{\spaceskip=\fontdimen2\font plus
\BIBentryALTinterwordstretchfactor\fontdimen3\font minus \fontdimen4\font\relax}
\providecommand{\BIBforeignlanguage}[2]{{%
\expandafter\ifx\csname l@#1\endcsname\relax
\typeout{** WARNING: IEEEtran.bst: No hyphenation pattern has been}%
\typeout{** loaded for the language `#1'. Using the pattern for}%
\typeout{** the default language instead.}%
\else
\language=\csname l@#1\endcsname
\fi
#2}}
\providecommand{\BIBdecl}{\relax}
\BIBdecl

\bibitem{rs14030589}
J.~Wang, D.~Li, W.~Cao, X.~Lou, A.~Shi, and H.~Zhang, ``Remote sensing analysis of erosion in {Arctic} coastal areas of {Alaska} and eastern {Siberia},'' \emph{Remote Sensing}, vol.~14, no.~3, 2022.

\bibitem{barmpoutis2020}
P.~Barmpoutis, P.~Papaioannou, K.~Dimitropoulos, and N.~Grammalidis, ``A review on early forest fire detection systems using optical remote sensing,'' \emph{Sensors}, vol.~20, no.~22, 2020.

\bibitem{radix}
\BIBentryALTinterwordspacing
``Radix,'' accessed: 2023-04-28. [Online]. Available: \url{https://space.skyrocket.de/doc\_sdat/radix.htm}
\BIBentrySTDinterwordspacing

\bibitem{MarCO}
\BIBentryALTinterwordspacing
``{MarCO},'' accessed: 2023-04-28. [Online]. Available: \url{https://www.jpl.nasa.gov/missions/mars-cube-one-marco}
\BIBentrySTDinterwordspacing

\bibitem{cubesat2020}
\BIBentryALTinterwordspacing
\emph{{CubeSat} design specification}, The CubeSat Program, Cal Poly SLO, 2022, rev. 14. [Online]. Available: \url{https://www.cubesat.org/cubesatinfo}
\BIBentrySTDinterwordspacing

\bibitem{nogales2018}
C.~Nogales, B.~Grim, M.~Kamstra, B.~Campbell, A.~Ewing, R.~Hance, J.~Griffin, and S.~Parke, ``{MakerSat-0}: {3D}-printed polymer degradation first data from orbit,'' 08 2018.

\bibitem{sentinel-2-user-handbook}
\BIBentryALTinterwordspacing
{European Space Agency}, ``Sentinel-2 user handbook,'' 2015. [Online]. Available: \url{https://sentinel.esa.int/documents/247904/685211/sentinel-2\_user\_handbook}
\BIBentrySTDinterwordspacing

\bibitem{6408177}
V.~Kostina and S.~Verdú, ``Lossy joint source-channel coding in the finite blocklength regime,'' \emph{IEEE Transactions on Information Theory}, vol.~59, no.~5, 2013.

\bibitem{cover1991elements}
T.~M. Cover and J.~A. Thomas, \emph{Elements of Information Theory}.\hskip 1em plus 0.5em minus 0.4em\relax Wiley-Interscience, 1991.

\bibitem{9838671}
T.-Y. Tung, D.~B. Kurka, M.~Jankowski, and D.~Gündüz, ``Deepjscc-q: Channel input constrained deep joint source-channel coding,'' in \emph{ICC 2022 - IEEE International Conference on Communications}, 2022.

\bibitem{Bourtsoulatze2019}
E.~Bourtsoulatze, D.~B. Kurka, and D.~Gunduz, ``Deep joint source-channel coding for wireless image transmission,'' \emph{{IEEE} Transactions on Cognitive Communications and Networking}, vol.~5, no.~3, sep 2019.

\bibitem{satjscc}
O.~Kondrateva, S.~Dietzel, and B.~Scheuermann, ``Joint source-and-channel coding for small satellite applications,'' in \emph{2023 IEEE 48th Conference on Local Computer Networks (LCN)}, 2023.

\bibitem{fontan2001}
F.~Fontan, M.~Vazquez-Castro, C.~Cabado, J.~Garcia, and E.~Kubista, ``Statistical modeling of the lms channel,'' \emph{IEEE Transactions on Vehicular Technology}, vol.~50, no.~6, 2001.

\bibitem{gallager1968information}
R.~Gallager, \emph{Information Theory and Reliable Communication}, ser. Courses and lectures.\hskip 1em plus 0.5em minus 0.4em\relax Wiley, 1968.

\bibitem{1614076}
Y.~Zhong, F.~Alajaji, and L.~Campbell, ``On the joint source-channel coding error exponent for discrete memoryless systems,'' \emph{IEEE Transactions on Information Theory}, vol.~52, no.~4, 2006.

\bibitem{4557472}
Y.~Zhong, F.~Alajaji, and L.~L. Campbell, ``Error exponents for asymmetric two-user discrete memoryless source-channel systems,'' in \emph{2007 IEEE International Symposium on Information Theory}, 2007.

\bibitem{Wei2004}
W.~Yu, Z.~Sahinoglu, and A.~Vetro, ``Energy efficient jpeg 2000 image transmission over wireless sensor networks,'' in \emph{IEEE Global Telecommunications Conference, 2004. GLOBECOM '04.}, vol.~5, 2004.

\bibitem{4205066}
Q.~Xu, V.~Stankovic, and Z.~Xiong, ``Distributed joint source-channel coding of video using raptor codes,'' \emph{IEEE Journal on Selected Areas in Communications}, vol.~25, no.~4, 2007.

\bibitem{toderici2016variable}
G.~Toderici, S.~M. O'Malley, S.~J. Hwang, D.~Vincent, D.~Minnen, S.~Baluja, M.~Covell, and R.~Sukthankar, ``Variable rate image compression with recurrent neural networks,'' 2016.

\bibitem{ballé2017endtoend}
J.~Ballé, V.~Laparra, and E.~P. Simoncelli, ``End-to-end optimized image compression,'' 2017.

\bibitem{Hu2022}
Y.~Hu, W.~Yang, Z.~Ma, and J.~Liu, ``Learning end-to-end lossy image compression: A benchmark,'' \emph{IEEE Transactions on Pattern Analysis and Machine Intelligence}, vol.~44, no.~8, 2022.

\bibitem{8054694}
T.~O’Shea and J.~Hoydis, ``An introduction to deep learning for the physical layer,'' \emph{IEEE Transactions on Cognitive Communications and Networking}, vol.~3, no.~4, 2017.

\bibitem{Xuan2021}
Z.~Xuan and K.~Narayanan, ``Deep joint source-channel coding for transmission of correlated sources over awgn channels,'' in \emph{ICC 2021 - IEEE International Conference on Communications}, 2021.

\bibitem{Kurka2021}
D.~B. Kurka and D.~Gündüz, ``Bandwidth-agile image transmission with deep joint source-channel coding,'' \emph{IEEE Transactions on Wireless Communications}, vol.~20, no.~12, 2021.

\bibitem{yang2021_2}
M.~Yang and H.-S. Kim, ``Deep joint source-channel coding for wireless image transmission with adaptive rate control,'' 2021.

\bibitem{9438648}
J.~Xu, B.~Ai, W.~Chen, A.~Yang, P.~Sun, and M.~Rodrigues, ``Wireless image transmission using deep source channel coding with attention modules,'' \emph{IEEE Transactions on Circuits and Systems for Video Technology}, vol.~32, no.~4, 2022.

\bibitem{nakajima2012}
A.~Nakajima, N.~Sako, M.~Kamemura, Y.~Wakayama, A.~Fukuzawa, H.~Sugiyama, and N.~Okad, ``Shindaisat: A visible light communication experimental micro-satellite,'' in \emph{Proc. Int. Conf. on Space Optical Sys. and App., (ICSOS)}, oct 2012.

\bibitem{welle2018}
R.~Welle, S.~Janson, D.~Rowen, and T.~Rose, ``Cubesat-scale high-speed laser downlinks,'' in \emph{Proc. of the 13th Reinventing Space Conf.}, 2018.

\bibitem{1623307}
C.~Loo, ``A statistical model for a land mobile satellite link,'' \emph{IEEE Transactions on Vehicular Technology}, vol.~34, no.~3, 1985.

\bibitem{Corazza1994ASM}
G.~E. Corazza and F.~Vatalaro, ``A statistical model for land mobile satellite channels and its application to nongeostationary orbit systems,'' \emph{IEEE Transactions on Vehicular Technology}, vol.~43, 1994.

\bibitem{596315}
S.-H. Hwang, K.-J. Kim, J.-Y. Ahn, and K.-C. Whang, ``A channel model for nongeostationary orbiting satellite system,'' in \emph{1997 IEEE 47th Vehicular Technology Conference. Technology in Motion}, vol.~1, 1997.

\bibitem{661055}
M.~Patzold, Y.~Li, and F.~Laue, ``A study of a land mobile satellite channel model with asymmetrical doppler power spectrum and lognormally distributed line-of-sight component,'' \emph{IEEE Transactions on Vehicular Technology}, vol.~47, no.~1, 1998.

\bibitem{966585}
F.~Fontan, M.~Vazquez-Castro, C.~Cabado, J.~Garcia, and E.~Kubista, ``Statistical modeling of the lms channel,'' \emph{IEEE Transactions on Vehicular Technology}, vol.~50, no.~6, 2001.

\bibitem{4151152}
S.~Scalise, C.~Alasseur, L.~Husson, and H.~Ernst, ``Sat04-2: Accurate and novel modeling of the land mobile satellite channel using reversible jump markov chain monte carlo technique,'' in \emph{IEEE Globecom 2006}, 2006.

\bibitem{7779114}
V.~Nikolaidis, N.~Moraitis, and A.~G. Kanatas, ``Dual-polarized narrowband mimo lms channel measurements in urban environments,'' \emph{IEEE Transactions on Antennas and Propagation}, vol.~65, no.~2, 2017.

\bibitem{8693582}
J.~J. Lopez-Salamanca, L.~O. Seman, M.~D. Berejuck, and E.~A. Bezerra, ``Finite-state markov chains channel model for cubesats communication uplink,'' \emph{IEEE Transactions on Aerospace and Electronic Systems}, vol.~56, no.~1, 2020.

\bibitem{9079470}
N.~Saeed, A.~Elzanaty, H.~Almorad, H.~Dahrouj, T.~Y. Al-Naffouri, and M.-S. Alouini, ``Cubesat communications: Recent advances and future challenges,'' \emph{IEEE Communications Surveys \& Tutorials}, vol.~22, no.~3, 2020.

\bibitem{sentinel2}
\BIBentryALTinterwordspacing
``{Sentinel2},'' accessed: 2023-04-28. [Online]. Available: \url{https://sentinel.esa.int/web/sentinel/missions/sentinel-2}
\BIBentrySTDinterwordspacing

\bibitem{8556744}
A.~P. Arechiga, A.~J. Michaels, and J.~T. Black, ``Onboard image processing for small satellites,'' in \emph{NAECON 2018 - IEEE National Aerospace and Electronics Conference}, 2018.

\bibitem{DBLP:journals/corr/SimonyanZ14a}
K.~Simonyan and A.~Zisserman, ``Very deep convolutional networks for large-scale image recognition,'' in \emph{3rd International Conference on Learning Representations, {ICLR} 2015, San Diego, CA, USA, May 7-9, 2015, Conference Track Proceedings}, Y.~Bengio and Y.~LeCun, Eds., 2015.

\bibitem{7780459}
K.~He, X.~Zhang, S.~Ren, and J.~Sun, ``Deep residual learning for image recognition,'' in \emph{2016 IEEE Conference on Computer Vision and Pattern Recognition (CVPR)}, 2016.

\bibitem{9884906}
E.~Dunkel, J.~Swope, Z.~Towfic, S.~Chien, D.~Russell, J.~Sauvageau, D.~Sheldon, J.~Romero-Cañas, J.~L. Espinosa-Aranda, L.~Buckley, E.~Hervas-Martin, M.~Fernandez, and C.~Knox, ``Benchmarking deep learning inference of remote sensing imagery on the qualcomm snapdragon and intel movidius myriad x processors onboard the international space station,'' in \emph{IGARSS 2022 - 2022 IEEE International Geoscience and Remote Sensing Symposium}, 2022.

\bibitem{wireless-attention-modules}
J.~Xu, B.~Ai, W.~Chen, A.~Yang, P.~Sun, and M.~Rodrigues, ``Wireless image transmission using deep source channel coding with attention modules,'' \emph{IEEE Transactions on Circuits and Systems for Video Technology}, vol.~32, no.~4, 2022.

\bibitem{resnet}
K.~He, X.~Zhang, S.~Ren, and J.~Sun, ``Deep residual learning for image recognition,'' \emph{arXiv preprint 1512.03385}, 2015.

\bibitem{channel-params}
H.~Smith, S.~K. Barton, J.~G. Gardiner, and M.~Sforza, ``Characterization of the land mobile-satellite {(LMS)} channel at {L} and {S} bands: Narrowband measurements,'' 1992, {ESA AOPs 104 433/114 473}.

\bibitem{DBLP:journals/ijscn/Perez-FontanMMPMMR08}
F.~P{\'{e}}rez{-}Font{\'{a}}n, A.~Mayo, D.~Marote, R.~Prieto{-}Cerdeira, P.~Mari{\~{n}}o, F.~Machado, and N.~Riera, ``Review of generative models for the narrowband land mobile satellite propagation channel,'' \emph{Int. J. Satell. Commun. Netw.}, vol.~26, no.~4, 2008.

\bibitem{7506756}
O.~Popescu, J.~S. Harris, and D.~C. Popescu, ``Designing the communication sub-system for nanosatellite cubesat missions: Operational and implementation perspectives,'' in \emph{SoutheastCon 2016}, 2016.

\bibitem{sumbul2019bigearthnet}
G.~Sumbul, M.~Charfuelan, B.~Demir, and V.~Markl, ``{BigEarthNet}: A large-scale benchmark archive for remote sensing image understanding,'' in \emph{IEEE International Geoscience and Remote Sensing Symposium}, Yokohama, Japan, 2019.

\bibitem{Sumbul2021}
G.~Sumbul, A.~de~Wall, T.~Kreuziger, F.~Marcelino, H.~Costa, P.~Benevides, M.~Caetano, B.~Demir, and V.~Markl, ``{BigEarthNet}-{MM}: A large-scale, multimodal, multilabel benchmark archive for remote sensing image classification and retrieval [software and data sets],'' \emph{{IEEE} Geoscience and Remote Sensing Magazine}, vol.~9, no.~3, Sep. 2021.

\bibitem{keras}
\BIBentryALTinterwordspacing
``{Keras}.'' [Online]. Available: \url{https://keras.io/}
\BIBentrySTDinterwordspacing

\bibitem{tensorflow}
\BIBentryALTinterwordspacing
``{Tensorflow}.'' [Online]. Available: \url{https://www.tensorflow.org/}
\BIBentrySTDinterwordspacing

\end{thebibliography}

\end{document}